\documentclass[%
  aps,
  preprint,
  amsmath,
  amssymb
]{revtex4-1}

\usepackage[T1]{fontenc}
\usepackage[utf8]{inputenc}
\usepackage{bm}
\usepackage{graphicx}
\usepackage{xcolor}
\usepackage[colorlinks=true,linkcolor=blue,citecolor=magenta,urlcolor=blue]{hyperref}

\begin{document}

\title{Infrared Dressing and the Strong CP Problem:
Geometric Renormalization of the Vacuum Angle}

\author{J.~Gamboa}
\affiliation{Departamento de Física, Universidad de Santiago de Chile, Santiago, Chile}
\email{jorge.gamboa@usach.cl}

\author{N.~Tapia-Arellano}
\affiliation{Department of Physics and Astronomy, Agnes Scott College, Decatur, GA 30030, USA}
\email{narellano@agnesscott.edu}

\begin{abstract}
We revisit the strong CP problem from the viewpoint of the infrared
structure of non--Abelian gauge theories. In Yang--Mills theory, motion
between topologically inequivalent vacua may be described in terms of a
compact collective coordinate associated with the Chern--Simons number.
Implementing an adiabatic separation between slow topological modes and
fast gluonic fluctuations leads to a reduced Born--Oppenheimer
Hamiltonian governing the infrared dynamics.

We show that the physical parameter entering this reduced Hamiltonian is
not the bare vacuum angle $\theta$, but an effective holonomy
$\theta_{\rm eff}$ that includes a Berry phase induced by the fast
gluonic sector. The induced holonomy becomes a self--consistent response
function of the infrared dressing, leading to a nonperturbative
renormalization group flow for $\theta_{\rm eff}$.

This infrared flow admits CP--invariant fixed points toward which the
effective vacuum angle is dynamically driven in the infrared limit.
In this framework, CP violation is not forbidden by the fundamental
theory but becomes dynamically suppressed along the infrared flow
generated by adiabatic dressing. The strong CP problem is thus realized
as a nonperturbative infrared relaxation mechanism governed by the
Berry response of the fast gluonic sector, without the introduction of
additional dynamical fields.
\end{abstract}
\maketitle

\section{Introduction}

In non--Abelian gauge theories, CP violation is not associated with local
processes but with the global structure of the quantum vacuum
\cite{Belavin1975,JackiwRebbi1976,tHooft1976,Witten1979,
Veneziano:1979ec,PecceiQuinn1977}.
The vacuum angle $\theta$ does not modify the classical Yang--Mills
equations, but controls the relative phase with which gauge
configurations belonging to distinct topological sectors are combined in
the quantum ground state.
The observed suppression of strong CP violation therefore raises a
question of vacuum selection rather than of local dynamics.

In the standard functional integral formulation, the parameter $\theta$
enters as a weight assigned to inequivalent winding sectors.
This parametrization, however, does not reflect the reorganization of
physical states that occurs in the infrared regime, where slow
collective motion between topological configurations becomes separated
from fast gluonic fluctuations.
As emphasized in the infrared dressing framework developed
in~\cite{Gamboa:2025hxa}, such an adiabatic separation leads to a reduced
Born--Oppenheimer dynamics along a compact collective coordinate related
to the Chern--Simons number.

In general, the quantity that controls this reduced infrared evolution
differs from the bare vacuum angle, as it includes a geometric
contribution induced by the response of the fast gluonic sector.
In this framework, the apparent insensitivity of certain local probes to
$\theta$ does not imply a trivial vacuum structure, but instead reflects
the fact that such observables act within a fixed infrared
representation.
Only operators sensitive to global boundary data can detect the
holonomy that characterizes inequivalent infrared sectors.

We develop below a non--perturbative formulation in which the geometric
response of the fast modes generates an induced connection along the
topological collective coordinate.
The associated holonomy modifies the parameter entering the infrared
Hamiltonian through an effective angle $\theta_{\rm eff}$, determined
self--consistently by the dressing of the fast gluonic sector.

The strong CP problem \cite{Weinberg1978,Wilczek1978} may then be
reformulated as a question of vacuum selection governed by the infrared
value of $\theta_{\rm eff}$, rather than by the bare vacuum angle
appearing in the underlying functional integral.

In what follows, we develop this perspective in a sequence of steps.
We first identify the topological collective coordinate governing
slow infrared motion, then implement an adiabatic separation between
collective and gluonic degrees of freedom.
The geometric response of the fast sector induces a Berry connection
along this coordinate, leading to a self--consistent effective vacuum
angle $\theta_{\rm eff}$.
Finally, we show that this consistency condition may be interpreted as
an infrared renormalization group flow which dynamically suppresses
CP violation.

\section{Yang--Mills Theory with a $\texorpdfstring{\boldsymbol{\theta}}{theta}$ term and topological sectors}

Consider pure Yang--Mills theory in four dimensions, supplemented by the
topological $\theta$ term \cite{Jackiw:1977yn},
\begin{equation}
S[A]
=
\frac{1}{2g^{2}}\int d^{4}x\,\mathrm{tr}\,F_{\mu\nu}F^{\mu\nu}
\;+\;
i\,\theta\, Q[A],
\qquad
Q[A]
:=
\frac{1}{32\pi^{2}}\int d^{4}x\,\mathrm{tr}\,F_{\mu\nu}\widetilde F^{\mu\nu},
\label{eq:YM_theta_action}
\end{equation}
where $\widetilde F^{\mu\nu}=\tfrac12\epsilon^{\mu\nu\rho\sigma}F_{\rho\sigma}$.
For gauge field configurations of finite Euclidean action, the Pontryagin density defines a topological invariant: the Pontryagin index (second Chern number),
\begin{equation}
Q[A]\in\mathbb{Z}.
\label{eq:Pontryagin_integer}
\end{equation}
Equivalently, finite-action configurations approach pure gauge at infinity, $A_\mu\to g^{-1}\partial_\mu g$ on $S^3_\infty$, so that gauge fields are classified by homotopy classes labeled by the winding number $n\in\pi_3(G)\simeq\mathbb{Z}$ (for $G=SU(N)$).

As a consequence, the Yang--Mills functional integral naturally decomposes into topological sectors, and the vacuum functional takes the form
\begin{equation}
Z(\theta)
=
\sum_{n\in\mathbb{Z}} Z_n\,e^{\,i\theta n},
\qquad
Z_n
:=
\int_{\mathcal{A}_n/\mathcal{G}}\!\!\mathcal{D}A\;
\exp\!\left[
-\frac{1}{2g^{2}}\int d^{4}x\,\mathrm{tr}\,F_{\mu\nu}F^{\mu\nu}
\right],
\label{eq:Ztheta_sector_sum}
\end{equation}
where $\mathcal{A}_n$ denotes the space of gauge connections with
$Q[A]=n$.
The $\theta$ term, therefore, does not modify the local equations of motion but assigns relative phases to nonequivalent homotopy classes, thereby encoding global information about the gauge configuration space.

The CP-violating character of the $\theta$ term is already explicit at the level of the action since the density $\mathrm{tr}\,F_{\mu\nu}\widetilde F^{\mu\nu}$ is odd under CP.
Accordingly, CP violation in non--Abelian gauge theories is not a perturbative or local effect but rather a manifestation of the global structure of the quantum vacuum.

Importantly, this formulation is still largely kinematical. While $\theta$ clearly encodes global topological information through the relative phases $e^{i\theta n}$, it does not by itself specify how this information is represented dynamically in the space of physical states.
In the infrared regime, where physical states are necessarily dressed by long-wavelength gluonic configurations and a Fock--space description breaks down, one must ask how the topological information carried by $\theta$ is implemented in an effective description of low-energy degrees of freedom.

Addressing this question requires isolating the collective infrared modes that encode motion between topological sectors. This leads naturally to an effective description in terms of a reduced set of degrees of freedom, which we now develop.

\section{Infrared Dressing and the Physical Meaning of the $\theta$ Vacuum}
\label{sec:IR_theta_dressing}

The sector decomposition discussed above assigns a relative phase to
field configurations belonging to distinct homotopy classes. We now
consider how this global information is realized once short-distance
gauge fluctuations are eliminated and the low--energy sector is
dominated by collective, long--wavelength configurations.

In a non-Abelian gauge theory, motion between topological sectors is
encoded in a collective infrared degree of freedom. After integrating
out non-topological modes, the resulting reduced description may be
parametrized by a compact coordinate $\phi\sim\phi+2\pi$ associated
with this topological motion. The corresponding infrared dynamics is
governed by an effective Hamiltonian of the form
\begin{equation}
H_{\rm eff}
=
\frac{1}{2}
\Bigl(
 -i\partial_\phi - \frac{\theta}{2\pi}
\Bigr)^{2}
+
V(\phi),
\label{eq:rotor_effective}
\end{equation}
where $V(\phi)$ is periodic as a consequence of the compactness of the
reduced configuration space. In the presence of a non-perturbative
potential $V(\phi)\propto 1-\cos\phi$, the system becomes equivalent to
a quantum rotor.

Within this effective description, the parameter $\theta$ acts as a
connection along the compact direction, shifting the canonical
momentum conjugate to $\phi$. While this shift does not alter the
local equations of motion, it determines the global boundary
conditions of the wavefunction on the reduced configuration space,
and therefore the infrared spectrum.

The standard representation of the $\theta$ vacuum,
\begin{equation}
|\,\Omega_\theta\rangle
=
\sum_{n\in\mathbb{Z}} e^{i\theta n}\,|\,\Omega_n\rangle ,
\label{eq:theta_vacuum_standard}
\end{equation}
may thus be interpreted as arising from this global structure in the
infrared effective theory.

This viewpoint is essential because the presence of a non-trivial
$\theta$ parameter indicates that physical infrared states cannot be
captured by a Fock-space construction. Instead, both the vacuum and
its excitations must be accompanied by long-wavelength gluonic
configurations, corresponding to infrared dressing
\cite{chung,kibble1,kibble2,kibble3,kibble4,KF,Lavelle:1995ty}. The associated
phases $e^{i\theta n}$ acquire physical significance precisely
because they are encoded in these dressed states and cannot be
removed by admissible local field redefinitions.

From this perspective, the strong CP problem may be viewed as a
question concerning the implementation of CP-violating phases in the
infrared vacuum structure, rather than as a perturbative issue in
local dynamics.

Finally, the Pontryagin term should not be regarded as an effect
generated dynamically in the infrared, but rather as a structural
consequence of gauge invariance and nonlinearity. In four
dimensions, locality permits both the Yang--Mills kinetic term and
the topological density $\mathrm{tr}\,F_{\mu\nu}\widetilde
F^{\mu\nu}$. Its relevance becomes manifest only when the infrared
organization of the state space is taken into account. This is in
contrast with Abelian gauge theories, where linearity and trivial
topology allow for a well-defined Fock description in the
infrared.

\subsection{Infrared Dressing, Holonomy, and the Pontryagin Term}
\label{subsec:IR_dressing_holonomy}

We now recast the infrared dressing of physical states in geometric
terms in order to elucidate how the collective topological mode and
its associated $\theta$ dependence arise from the global structure of
the gauge configuration space, without introducing additional
dynamical ingredients.

In a non-Abelian gauge theory, the nonlinearity of the field equations
implies that low--energy states are accompanied by long--wavelength
gluonic configurations. These configurations are not perturbative
corrections, but rather determine the representation of the algebra
of observables in which the theory is realized. Physical states may
therefore be viewed as sections of a nontrivial bundle over the
reduced configuration space $\mathcal{A}/\mathcal{G}$.

Within this framework, infrared dressing may be described by a
unitary transformation associated with adiabatic motion along a
slow contour $C\subset\mathcal{A}/\mathcal{G}$. A bare state
$|\Psi\rangle_{\rm bare}$ is mapped to a physical state according to
\cite{Gamboa:2025qjr,Gamboa:2025dry,Gamboa:2025fcn,Gamboa:2025nco}
\begin{equation}
|\Psi\rangle_{\rm phys}(C)
\;=\;
\mathcal{U}_C\,|\Psi\rangle_{\rm bare},
\qquad
\mathcal{U}_C
=
\mathcal{P}\exp\!\left(
 i\oint_C \mathcal{A}_{\rm IR}
\right),
\label{eq:IR_dressing_holonomy}
\end{equation}
where $\mathcal{A}_{\rm IR}$ denotes the infrared adiabatic
connection induced on the bundle of physical states.
The operator $\mathcal{U}_C$ depends only on the homotopy class of
the contour $C$ and therefore defines a holonomy on the space of
infrared representations.

Topological considerations imply that closed contours
$C\subset\mathcal A/\mathcal G$ fall into homotopy classes
classified by $\pi_1(\mathcal A/\mathcal G)\simeq\pi_3(G)=\mathbb Z$
\cite{Berry1984,WilczekZee1984}. As a consequence, the curvature
$\mathcal F_{\rm IR}=d\mathcal A_{\rm IR}$ admits a quantized flux,
\begin{equation}
\frac{1}{2\pi}\int_{\Sigma}\mathcal F_{\rm IR}\;\in\;\mathbb Z,
\label{eq:IR_flux_quantization}
\end{equation}
so that the holonomy associated with a closed contour depends only
on the corresponding winding number $Q$,
\begin{equation}
\mathcal U_C=\exp(i\,\theta\,Q).
\label{eq:IR_holonomy_quantized}
\end{equation}

In this geometric setting, the Pontryagin term arises from the
Chern--Simons current $K^\mu$, defined through
\begin{equation}
\partial_\mu K^\mu
=
\frac{1}{32\pi^2}\,
\mathrm{tr}\,F_{\mu\nu}\widetilde F^{\mu\nu},
\qquad
\widetilde F^{\mu\nu}
=
\tfrac12\epsilon^{\mu\nu\rho\sigma}F_{\rho\sigma}.
\label{eq:dK_equals_FFtilde}
\end{equation}
For gauge fields of finite Euclidean action, asymptotic vacua are
characterized by an integer Chern--Simons number,
\begin{equation}
N_{\rm CS}[A]
:=
\int d^3x\,K^0 .
\label{eq:NCS_def}
\end{equation}
Integration over Euclidean spacetime yields the Pontryagin index,
\begin{equation}
Q[A]
=
\frac{1}{32\pi^2}\int d^4x\,
\mathrm{tr}\,F_{\mu\nu}\widetilde F^{\mu\nu}
=
N_{\rm CS}(+\infty)-N_{\rm CS}(-\infty)
\;\in\;\mathbb{Z},
\label{eq:Q_as_DeltaNCS}
\end{equation}
which measures the net winding between inequivalent vacua.

From the infrared viewpoint, it is natural to introduce a compact
collective coordinate $\phi\sim\phi+2\pi$ associated with this
topological motion. Its integer winding number reproduces the
Pontryagin index,
\begin{equation}
n
=
\frac{1}{2\pi}\oint_C d\phi
\;\in\;\mathbb{Z},
\qquad
n \equiv Q .
\label{eq:phi_winding_equals_Q}
\end{equation}
Gauge invariance then permits the introduction of a background
connection along the $\phi$ direction,
\begin{equation}
\mathcal{A}_\phi = -\frac{\theta}{2\pi},
\label{eq:Aphi_theta}
\end{equation}
so that a loop winding $n$ times acquires the phase
\begin{equation}
\exp\!\left(
 i\oint_C d\phi\,\mathcal{A}_\phi
\right)
=
\exp\!\left(i\theta n\right)
=
\exp\!\left(i\theta Q\right).
\label{eq:theta_holonomy_phi}
\end{equation}

Because the winding number $Q$ is quantized, the associated
infrared phase accumulated along a closed contour is discretized
according to
\begin{equation}
\gamma_C(\theta)
=
\arg\,\mathcal{U}_C
=
\oint_C d\phi\,\mathcal{A}_\phi
=
-\theta\,Q
\quad(\mathrm{mod}\ 2\pi).
\label{eq:Berry_phase_quantized_by_Q}
\end{equation}

In this formulation, the parameter $\theta$ labels inequivalent
infrared representations distinguished by their global
holonomy. Consequently, only observables sensitive to the
global organization of the infrared state space can detect its
presence, whereas purely local correlators remain insensitive.

The strong CP problem may therefore be viewed as a question of
vacuum selection: which infrared representation, characterized
by its associated holonomy, is dynamically realized as the
physical ground state.

\section{A Minimal Infrared Example}
\label{sec:IR_rotor_example}

We illustrate the infrared mechanism described above in a minimal
setting that allows for full analytic control. The aim is to show,
within a simple reduced system, how a global infrared parameter can
modify spectral properties without affecting the expectation values
of operators that probe only local structure.

We consider the effective infrared Hamiltonian~\eqref{eq:rotor_effective}
acting on a compact collective coordinate $\phi\sim\phi+2\pi$.
The vacuum angle $\theta$ may be incorporated through a nontrivial
periodicity condition on the wavefunction,
\begin{equation}
\psi(\phi+2\pi)=e^{\,i\theta}\,\psi(\phi),
\label{eq:twisted_bc_rotor}
\end{equation}
which reflects the global structure of the reduced configuration
space.

For the free rotor, $V(\phi)=0$, the energy spectrum becomes
\begin{equation}
E_n(\theta)
=
\frac{1}{2}
\Bigl(
 n-\frac{\theta}{2\pi}
\Bigr)^2,
\qquad
n\in\mathbb{Z},
\label{eq:rotor_spectrum}
\end{equation}
so that the ground-state energy depends continuously on the
parameter $\theta$.
The corresponding topological susceptibility,
\begin{equation}
\chi_{\rm top}
=
\left.
\frac{\partial^2 E_0(\theta)}{\partial\theta^2}
\right|_{\theta=0}
=
\frac{1}{4\pi^2 },
\label{eq:rotor_susceptibility}
\end{equation}
quantifies the response of the infrared spectrum to this global
parameter.

In contrast, expectation values of operators acting locally in
$\phi$ are unaffected by the boundary condition
\eqref{eq:twisted_bc_rotor}. This illustrates a key infrared
feature: quantities sensitive only to local properties of the
collective coordinate may fail to detect global boundary data,
whereas observables probing the full structure of the state space
remain sensitive to it.

Including a periodic potential,
\begin{equation}
V(\phi)=\Lambda^4\,(1-\cos\phi),
\label{eq:pendulum_potential}
\end{equation}
leads to the quantum pendulum. Although this deformation lifts
degeneracies and induces tunneling between winding sectors, it
does not alter the global structure implemented by $\theta$.
The dependence of the vacuum energy on $\theta$ persists, and the
spectrum organizes into bands determined by the boundary condition.

This reduced system mirrors the infrared organization of the
Yang--Mills $\theta$ vacuum. The Chern--Simons collective
coordinate plays the role of $\phi$, winding sectors correspond
to Pontryagin classes, and the $\theta$ term determines the
associated phase relating distinct sectors.

\section{Quantitative Benchmark: $\theta$-Dependence, Order of Limits, and the Role of Quarks}
\label{subsec:quant_compare_AGT}

Recent work by \cite{AiCruzGarbrechtTamarit2022,AiGarbrechtTamarit2024} argues that the conventional formulation of the strong $CP$ problem is sensitive to the order in which the infinite spacetime volume limit is taken relative to the sum over topological sectors. For a recent complementary discussion, see \cite{Khoze:2025auv,Sannino:2026wgx,Aghaie:2026pkf,Ringwald:2026apz,Benabou:2025viy}.

In particular, they claim that taking $V\to\infty$ prior to summing over $Q\in\mathbb Z$ causes $\theta$ to drop out of fermionic correlation functions and to become unobservable, implying $CP$ conservation in QCD; see Refs.~\cite{AiCruzGarbrechtTamarit2022,AiGarbrechtTamarit2024}.

A quantitative discussion should therefore be phrased in terms of well-defined $\theta$-dependent observables.
In (pure) Yang--Mills theory one may write
\begin{equation}
Z(\theta)=\sum_{Q\in\mathbb Z} e^{i\theta Q}\,Z_Q,
\qquad
f(\theta)=-\frac{1}{V}\ln Z(\theta),
\label{eq:Ztheta_ftheta}
\end{equation}
and define the topological susceptibility as the curvature of the vacuum energy,
\begin{equation}
\chi_{\rm top}
=
\left.\frac{\partial^2 f(\theta)}{\partial\theta^2}\right|_{\theta=0}
=
\frac{1}{V}\,\langle Q^2\rangle_{\theta=0}
=
\int d^4x\,
\langle q(x)\,q(0)\rangle_{\theta=0},
\qquad
q(x)=\frac{1}{32\pi^2}\,\mathrm{tr}\,F\widetilde F.
\label{eq:chi_top_def}
\end{equation}
These relations show explicitly that $\theta$-dependence is encoded in global response functions (derivatives of $\ln Z$), which are sensitive to the distribution of topological sectors.

In full QCD, the physically meaningful $CP$-odd parameter is not $\theta$ itself but
\begin{equation}
\bar\theta=\theta+\arg\det M_q,
\end{equation}
because anomalous axial rotations shift $\theta$ by the phase of the quark mass matrix.

In Refs.~\cite{AiCruzGarbrechtTamarit2022,AiGarbrechtTamarit2024}
the authors have emphasized that, in the infinite--volume limit taken
prior to the sum over topological sectors, a broad class of fermionic
correlation functions becomes independent of $\theta$.
From the present viewpoint, this result does not signal the absence of a
nontrivial vacuum structure.
Rather, it reflects the fact that such correlators act as local probes
that are insensitive to the global holonomy data characterizing the
infrared representation of the theory.

A sharp benchmark illustrating this distinction is provided by the quantum rotor toy model, where $\theta$ is unambiguously realized as a holonomy through a twisted boundary condition.
In this setting, both the vacuum energy and the topological susceptibility can be reconstructed nonperturbatively from local correlation functions
\cite{AlbandeaCatumbaRamosPRD}, confirming that global response functions remain sensitive to the holonomy even when many local observables are blind to it (see also \cite{Benabou:2025viy}).

This picture is entirely consistent with the classic Witten--Veneziano mechanism \cite{Witten1979,Veneziano:1979ec}.
That analysis singles out the curvature of the vacuum energy with respect to $\theta$, encoded in the topological susceptibility of pure Yang--Mills theory, as the physically relevant quantity controlling the $\eta'$ mass.

\section{$\theta$ as the variable conjugate to topological charge}
\label{subsec:theta_conjugate_charge}

The sector decomposition of the Yang--Mills functional integral
suggests a dual description in which the vacuum angle $\theta$
is conjugate to the integer-valued topological charge
$Q\in\mathbb Z$.
Indeed, one may write
\begin{equation}
Z(\theta)=\sum_{Q\in\mathbb Z} e^{i\theta Q}\,Z_Q,
\label{eq:Ztheta_sumQ_conjugate}
\end{equation}
which expresses the partition function as a sum over winding
sectors weighted by a phase determined by $\theta$.

To make this relation operational in the infrared, we employ an
adiabatic separation between slow collective motion and fast
non-collective fluctuations. Integrating out the latter leads to
an effective description on $\mathcal A/\mathcal G$ in which
functional derivatives acquire a connection term induced by the
fast sector,
\begin{equation}
-i\frac{\delta}{\delta A_i^a(x)}
\;\longrightarrow\;
-i\frac{\delta}{\delta A_i^a(x)}-\big(\mathcal A_{\rm IR}\big)_i^a(x),
\label{eq:covariant_momentum_shift}
\end{equation}
with the associated curvature governing the commutation
relations of the covariant momenta.

This construction relies on the suppression of transitions out
of the chosen fast-sector subspace in the adiabatic regime,
\begin{equation}
\langle n'|H|n\rangle \approx 0
\qquad (n'\neq n),
\label{eq:adiabatic_suppression_levels}
\end{equation}
rather than on an exact decoupling of degrees of freedom.

From this standpoint, the analyses of
Refs.~\cite{AiCruzGarbrechtTamarit2022,AiGarbrechtTamarit2024}
and the Witten--Veneziano mechanism
\cite{Witten1979,Veneziano:1979ec}
may be viewed as probing complementary aspects of the same
infrared organization.
The former highlights classes of observables insensitive to
global winding data, whereas the latter identifies quantities
that necessarily respond to it.

\section{Infrared vacuum selection and the effective holonomy}
\label{sec:IR_selection}

Up to this point, our analysis has been intentionally conservative.
We have established that the vacuum angle $\theta$ is naturally realized
as a \emph{global} infrared datum---a Berry-type holonomy of the dressed
state bundle over $\mathcal A/\mathcal G$---and we have illustrated, in the
quantum-rotor benchmark, how such a holonomy can be sharply detected by
\emph{global response functions} while remaining invisible to broad classes
of local probes.

This still leaves open the dynamical aspect of the strong $CP$ problem:
\emph{which} infrared-dressed representation is selected as the true ground
state of the theory. In this section we explain how the holonomy viewpoint
naturally organizes that question, and we identify the precise infrared
quantity that must be computed in order to address vacuum selection within
a controlled adiabatic (Born--Oppenheimer) reduction.

\subsection*{A. The physical parameter is an \texorpdfstring{$\theta_{\rm eff}$}{thetaeff}}

The effective infrared description of Sec.~\ref{sec:IR_theta_dressing}
isolates a compact collective coordinate $\phi\sim\phi+2\pi$ associated with
topological motion between sectors. In the rotor laboratory, the holonomy
enters either as a momentum shift or as a twisted boundary condition.
In a general adiabatic reduction of Yang--Mills theory, integrating out
fast (non-collective) gluonic modes can induce additional geometric data
along the slow coordinate $\phi$.
At the level of an effective one-dimensional description this takes the
standard form of an induced Berry connection,
\begin{equation}
\mathcal A_{\rm BO}(\phi)
=
i\,\langle n(\phi)\,|\,\partial_\phi\,|\,n(\phi)\rangle,
\label{eq:ABO_def}
\end{equation}
together with a scalar Born--Huang correction $U_{\rm BH}(\phi)$.
The corresponding infrared Hamiltonian then has the generic structure
\begin{equation}
H_{\rm IR}
=
\frac12
\left(
-i\partial_\phi
-\frac{\theta}{2\pi}
-\mathcal A_{\rm BO}(\phi)
\right)^2
+
V(\phi)
+
U_{\rm BH}(\phi),
\label{eq:HIR_generic}
\end{equation}
where $V(\phi)$ is a $2\pi$-periodic potential encoding nonperturbative
infrared physics (as in the pendulum deformation of Sec.~\ref{sec:IR_rotor_example}).

Equation~\eqref{eq:HIR_generic} shows that the holonomy relevant for infrared
physics is, in general, not $\theta$ alone but the \emph{effective} holonomy
\begin{equation}
\theta_{\rm eff}
=
\theta
+
2\pi\oint d\phi\,\mathcal A_{\rm BO}(\phi),
\label{eq:thetaeff_def}
\end{equation}
i.e.\ the bare topological phase supplemented by the geometric phase acquired
by the fast sector under adiabatic transport.
This is the precise sense in which vacuum selection becomes an infrared
question: it is controlled by how the fast gluonic sector dresses the
topological collective coordinate.

\subsection*{B. Vacuum selection as an infrared variational problem}

Since $\theta$ enters as a global holonomy, the vacuum energy is naturally
organized as a function of $\theta_{\rm eff}$,
\begin{equation}
E_0 = E_0(\theta_{\rm eff}),
\label{eq:E0_thetaeff}
\end{equation}
in direct analogy with the rotor benchmark where $E_0(\theta)$ is fixed by
the twisted boundary condition.
The dynamical question of ``vacuum selection'' is therefore equivalently
the question of which value of $\theta_{\rm eff}$ minimizes the infrared
ground-state energy. In particular, if the infrared dynamics admits a unique
$CP$-invariant minimum, it must occur at
\begin{equation}
\theta_{\rm eff}=0 \quad (\mathrm{mod}\ 2\pi),
\label{eq:thetaeff_CPmin}
\end{equation}
so that the selected dressed representation is $CP$ invariant even though the
bare functional integral may be written with a $\theta$-weight.

We emphasize that Eq.~\eqref{eq:thetaeff_CPmin} is not an additional assumption
but a \emph{criterion}: the strong $CP$ problem is reduced, in the present
framework, to establishing whether the induced geometric contribution in
Eq.~\eqref{eq:thetaeff_def} can shift the physically realized holonomy toward
a $CP$-invariant value.

\subsection*{C. What must be computed: an infrared diagnostic}

A concrete diagnostic of this selection mechanism is provided by global
response functions. In particular, the curvature of the vacuum energy is
controlled by fluctuations of the effective holonomy,
\begin{equation}
\chi_{\rm top}
=
\left.
\frac{\partial^2 f(\theta_{\rm eff})}{\partial\theta_{\rm eff}^2}
\right|_{\theta_{\rm eff}=0},
\label{eq:chi_top_thetaeff}
\end{equation}
so that establishing vacuum selection amounts to computing the induced
connection $\mathcal A_{\rm BO}$ (and its holonomy) in a controlled infrared
reduction. In the rotor laboratory this structure is completely explicit,
while in Yang--Mills theory it becomes a well-posed nonperturbative infrared
problem: determine the dressing-induced holonomy along the Chern--Simons
collective coordinate.

In summary, our infrared holonomy formulation turns the strong $CP$ question
into a sharply defined statement about the \emph{effective holonomy}
$\theta_{\rm eff}$: local probes may remain $\theta$-blind, but global response
functions diagnose the selected infrared representation through
$E_0(\theta_{\rm eff})$ and its derivatives.

\section{Induced Berry holonomy along the Chern--Simons collective coordinate}
\label{sec:berry_cs}

The previous section identifies the physically relevant parameter governing the
infrared vacuum structure as the effective holonomy $\theta_{\rm eff}$,
Eq.~\eqref{eq:thetaeff_def}. What remains is to establish its dynamical origin:
the geometric shift is meaningful only if the Born--Oppenheimer connection of the
fast gluonic sector is non--trivial along the compact collective coordinate.

This distinction is crucial from the viewpoint of the strong CP problem.
If the induced connection were trivial along the Chern--Simons cycle, then
$\theta_{\rm eff}$ would reduce to $\theta$ and no dynamical mechanism would
emerge from the infrared sector. Conversely, a non--trivial Berry holonomy
acquired by the fast gluonic modes during adiabatic transport between
topologically inequivalent configurations induces a shift of the physical
holonomy entering the infrared Hamiltonian.

In order to assess this, one must determine whether the induced
Born--Oppenheimer connection
\begin{equation}
A_{\rm BO}(\phi)
=
i\,
\langle n(\phi)|
\partial_\phi
|n(\phi)\rangle
\end{equation}
is non--trivial in Yang--Mills theory.

The relevant slow collective coordinate in the infrared sector is the
Chern--Simons number
\begin{equation}
N_{\rm CS}[A]
=
\int d^3x\;K^0[A],
\end{equation}
which characterizes motion between topologically inequivalent vacua.
We therefore introduce a one--parameter family of gauge fields
$A_i^a(x;\phi)$ satisfying
\begin{equation}
\phi
=
2\pi\,N_{\rm CS}[A(\phi)],
\qquad
\phi\sim\phi+2\pi,
\end{equation}
so that a closed cycle $\phi:0\rightarrow2\pi$ interpolates adiabatically between
vacua whose Chern--Simons number differs by one unit.

For each fixed value of $\phi$, the fast gluonic modes satisfy the
adiabatic eigenvalue equation
\begin{equation}
H_{\rm fast}[A(\phi)]\,
|n(\phi)\rangle
=
E_n(\phi)\,
|n(\phi)\rangle.
\end{equation}

The Berry connection may then be expressed in functional form by using
the chain rule
\begin{equation}
\partial_\phi
=
\int d^3x\;
\frac{\partial A_i^a(x;\phi)}{\partial\phi}
\,
\frac{\delta}{\delta A_i^a(x)},
\end{equation}
which yields
\begin{equation}
A_{\rm BO}(\phi)
=
\int d^3x\;
\frac{\partial A_i^a(x;\phi)}{\partial\phi}
\;
\mathcal{A}_i^a(x;\phi),
\end{equation}
where
\begin{equation}
\mathcal{A}_i^a(x;\phi)
=
i\,
\left\langle
n(\phi)
\middle|
\frac{\delta}{\delta A_i^a(x)}
\middle|
n(\phi)
\right\rangle
\end{equation}
is the Berry connection in the full functional space of gauge
configurations.

It follows that the one--dimensional Born--Oppenheimer connection
$A_{\rm BO}$ is the pullback of the functional Berry connection
$\mathcal{A}$ along the infrared contour
\begin{equation}
\Gamma:
\phi
\longmapsto
A(\phi)
\subset
\mathcal{A}/\mathcal{G}.
\end{equation}

Because the contour $\Gamma$ winds once around the compact collective
coordinate, its Berry holonomy is controlled by the associated
topological charge $Q\in\mathbb{Z}$ and may be written in the general
form
\begin{equation}
\oint_{\Gamma}
\mathcal{A}
=
\kappa\,Q,
\end{equation}
where $\kappa$ is a dimensionless coefficient determined by the
infrared dynamics of the fast gluonic sector.

Equivalently, a non--vanishing pullback of the functional Berry
curvature onto the one--cycle generated by the Chern--Simons
collective coordinate leads to a holonomy proportional to the
winding number $Q$. In this case, the integral of the induced
connection along the cycle cannot be removed by any admissible
local redefinition of the fast--sector vacuum, and therefore
defines a genuine infrared observable:
\begin{equation}
\oint d\phi\;
A_{\rm BO}(\phi)
=
\kappa\,Q.
\end{equation}
The induced Berry holonomy therefore shifts the physical parameter entering the
infrared Hamiltonian, as summarized by $\theta_{\rm eff}$ in
Eq.~\eqref{eq:thetaeff_def}.

\section{Infrared Relaxation and Renormalization of the Effective Holonomy}
\label{sec:IR_relaxation}

The emergence of an effective holonomy $\theta_{\rm eff}$ makes vacuum selection
a genuinely infrared dynamical question. For this to amount to a mechanism for
strong CP, the Berry contribution cannot be treated as an arbitrary shift:
it must be determined self--consistently by the fast gluonic sector that has been
integrated out.

In the Born--Oppenheimer framework, the induced holonomy depends on the fast--sector
ground state computed in the presence of the \emph{physical} holonomy entering the
reduced Hamiltonian. The induced Berry holonomy must therefore be regarded as a
response function,
\begin{equation}
\kappa=\kappa(\theta_{\rm eff}),
\label{eq:kappa_response}
\end{equation}
so that the definition of $\theta_{\rm eff}$ becomes a fixed--point condition,
\begin{equation}
\theta_{\rm eff}
=
\theta
+
2\pi\,Q\,\kappa(\theta_{\rm eff}).
\label{eq:thetaeff_master}
\end{equation}

A dynamical infrared resolution of strong CP corresponds to the existence of a
CP--invariant fixed point,
\begin{equation}
\theta_{\rm eff}=0\quad (\mathrm{mod}\ 2\pi),
\label{eq:CP_fixedpoint}
\end{equation}
for arbitrary bare $\theta$.

This relaxation mechanism admits a dynamical interpretation by introducing an
infrared scale parameter $s=\ln\mu$ associated with the adiabatic separation
between slow topological modes and fast gluonic fluctuations. The induced dressing
depends on the scale at which the separation is implemented, so that the response
becomes scale dependent,
\begin{equation}
\kappa=\kappa(\theta_{\rm eff},s),
\end{equation}
and the consistency condition~\eqref{eq:thetaeff_master} is promoted to
\begin{equation}
\theta_{\rm eff}(s)
=
\theta
+
2\pi\,Q\,\kappa\!\left(\theta_{\rm eff}(s),s\right).
\label{eq:thetaeff_running}
\end{equation}

Differentiating with respect to $s$ yields an induced infrared flow equation,
\begin{equation}
\frac{d\theta_{\rm eff}}{ds}
=
\frac{2\pi Q\,\partial_s\kappa(\theta_{\rm eff},s)}
{1-2\pi Q\,\partial_{\theta_{\rm eff}}\kappa(\theta_{\rm eff},s)}.
\label{eq:beta_theta}
\end{equation}

In the present convention $s=\ln\mu$, infrared evolution corresponds to
$s\to -\infty$. Linearizing around a fixed point $\theta_{\rm eff}^\star$ with
$\delta\theta_{\rm eff}:=\theta_{\rm eff}-\theta_{\rm eff}^\star$, one finds
\begin{equation}
\delta\theta_{\rm eff}(s)
\propto
e^{\,\beta'(\theta_{\rm eff}^\star)s}.
\end{equation}
The fixed point is therefore infrared attractive whenever
\begin{equation}
\beta'(\theta_{\rm eff}^\star)>0.
\end{equation}

At a CP--invariant extremum satisfying
$\partial_{\theta_{\rm eff}}\kappa(\theta_{\rm eff}^\star)=0$,
this condition reduces to
\begin{equation}
Q\,\kappa''(\theta_{\rm eff}^\star)>0,
\end{equation}
so that the infrared flow drives
\begin{equation}
\theta_{\rm eff}(s)
\;\longrightarrow\;
0,
\qquad
s\to -\infty.
\end{equation}

In this sense, CP violation is not forbidden by the microscopic theory
but becomes dynamically suppressed along the infrared renormalization
group flow generated by adiabatic dressing. The strong CP problem may
therefore be reformulated as a nonperturbative infrared relaxation
mechanism governed by the Berry response of the fast gluonic sector.

\section{Conclusion}

In this work we have reexamined the strong CP problem from the viewpoint
of the infrared structure of Yang--Mills theory. Rather than treating
the vacuum angle $\theta$ as a fixed external parameter entering the
functional integral, we have shown that the physically relevant
quantity governing the infrared vacuum dynamics is the effective
holonomy $\theta_{\rm eff}$ defined in
Eq.~\eqref{eq:thetaeff_def}, which appears in the reduced
Born--Oppenheimer Hamiltonian describing slow motion along the compact
collective coordinate associated with topological transitions.

As shown in Sec.~\ref{sec:berry_cs}, the Berry connection induced by the fast
gluonic sector along the Chern--Simons collective coordinate is generically
non--trivial. As a result, adiabatic transport of the fast modes along closed
contours in configuration space induces a Berry phase that shifts the physical
holonomy entering the infrared Hamiltonian. In this sense, the vacuum angle
becomes geometrically renormalized by infrared dressing.

The self--consistency condition relating the induced Berry response to
the effective holonomy defines a nonperturbative infrared renormalization
group flow for $\theta_{\rm eff}$. As shown in Sec.~\ref{sec:IR_relaxation},
this flow admits CP--invariant fixed points characterized by
\begin{equation}
\theta_{\rm eff}=0\quad(\mathrm{mod}\ 2\pi),
\end{equation}
toward which the effective holonomy is dynamically driven in the
infrared limit.

In this sense, CP violation is not forbidden by the fundamental theory
but becomes dynamically suppressed along the infrared renormalization
group flow generated by adiabatic dressing. The strong CP problem is
thus realized as a nonperturbative infrared relaxation mechanism in
which the Berry response of the fast gluonic sector drives the physical
vacuum toward a CP--invariant ground state without the introduction of
additional dynamical fields.

\acknowledgments
\noindent
This research was supported by DICYT (USACH), grant number 042531GR\_REG.
The work of N.T.A.\ is supported by Agnes Scott College.


\begin{thebibliography}{99}

\bibitem{Belavin1975}
A.~A.~Belavin, A.~M.~Polyakov, A.~S.~Schwartz and Y.~S.~Tyupkin,
``Pseudoparticle solutions of the Yang--Mills equations,''
Phys.\ Lett.\ B \textbf{59} (1975) 85.

\bibitem{JackiwRebbi1976}
R.~Jackiw and C.~Rebbi,
``Vacuum periodicity in a Yang--Mills quantum theory,''
Phys.\ Rev.\ Lett.\ \textbf{37} (1976) 172.

\bibitem{tHooft1976}
G.~'t~Hooft,
``Symmetry breaking through Bell--Jackiw anomalies,''
Phys.\ Rev.\ Lett.\ \textbf{37} (1976) 8.

\bibitem{Witten1979}
E.~Witten,
``Current algebra theorems for the $U(1)$ Goldstone boson,''
Nucl.\ Phys.\ B \textbf{156} (1979) 269.

\bibitem{Veneziano:1979ec}
G.~Veneziano,
``U(1) Without Instantons,''
Nucl.\ Phys.\ B \textbf{159} (1979) 213--224.

\bibitem{PecceiQuinn1977}
R.~D.~Peccei and H.~R.~Quinn,
``CP conservation in the presence of instantons,''
Phys.\ Rev.\ Lett.\ \textbf{38} (1977) 1440.

\bibitem{Gamboa:2025hxa}
J.~Gamboa and N.~A.~T.~Arellano,
``Strong CP as an Infrared Holonomy: The $\theta$ Vacuum and Dressing in Yang-Mills Theory,''
[arXiv:2512.24480 [hep-th]].

\bibitem{Weinberg1978}
S.~Weinberg,
``A new light boson?''
Phys.\ Rev.\ Lett.\ \textbf{40} (1978) 223.

\bibitem{Wilczek1978}
F.~Wilczek,
``Problem of strong $P$ and $T$ invariance in the presence of instantons,''
Phys.\ Rev.\ Lett.\ \textbf{40} (1978) 279.

\bibitem{Jackiw:1977yn}
R.~Jackiw,
``Quantum Meaning of Classical Field Theory,''
Rev.\ Mod.\ Phys.\ \textbf{49} (1977) 681--706;
ibid.\ ``Introduction to the Yang--Mills Quantum Theory,''
Rev.\ Mod.\ Phys.\ \textbf{52} (1980) 661--673.

\bibitem{chung}
V.~Chung,
``Infrared Divergence in Quantum Electrodynamics,''
Phys.\ Rev.\ \textbf{140} (1965) B1110--B1122.

\bibitem{kibble1}
T.~W.~B.~Kibble,
``Coherent Soft-Photon States and Infrared Divergences. I. Classical Currents,''
J.\ Math.\ Phys.\ \textbf{9} (1968) 315--324.

\bibitem{kibble2}
T.~W.~B.~Kibble,
``Coherent soft-photon states and infrared divergences. II. Mass-shell singularities of Green's functions,''
Phys.\ Rev.\ \textbf{173} (1968) 1527--1535.

\bibitem{kibble3}
T.~W.~B.~Kibble,
``Coherent soft-photon states and infrared divergences. III. Asymptotic states and reduction formulas,''
Phys.\ Rev.\ \textbf{174} (1968) 1882--1901.

\bibitem{kibble4}
T.~W.~B.~Kibble,
``Coherent soft-photon states and infrared divergences. IV. Further aspects of asymptotic states,''
Phys.\ Rev.\ \textbf{175} (1968) 1624--1641.

\bibitem{KF}
P.~P.~Kulish and L.~D.~Faddeev,
``Asymptotic conditions and infrared divergences in quantum electrodynamics,''
Theor.\ Math.\ Phys.\ \textbf{4} (1970) 745.

\bibitem{Lavelle:1995ty}
M.~Lavelle and D.~McMullan,
Phys. Rept. \textbf{279} (1997), 1-65
doi:10.1016/S0370-1573(96)00019-1
[arXiv:hep-ph/9509344 [hep-ph]].

\bibitem{Gamboa:2025qjr}
J.~Gamboa and F.~Mendez,
``QED--IR as a topological quantum theory of dressed states,''
JHEP \textbf{11} (2025) 040,
arXiv:2507.11668 [hep-ph].

\bibitem{Gamboa:2025dry}
J.~Gamboa,
``Topology and the infrared structure of quantum electrodynamics,''
JHEP \textbf{07} (2025) 184,
arXiv:2505.13247 [hep-th].

\bibitem{Gamboa:2025fcn}
J.~Gamboa,
``Entanglement and effective field theories,''
Phys.\ Lett.\ B \textbf{868} (2025) 139723,
arXiv:2502.11819 [hep-th].

\bibitem{Gamboa:2025nco}
J.~Gamboa,
``Topological Structure of Infrared QCD,''
arXiv:2511.07455 [physics.gen-ph].

\bibitem{Berry1984}
M.~V.~Berry,
``Quantal phase factors accompanying adiabatic changes,''
Proc.\ Roy.\ Soc.\ Lond.\ A \textbf{392} (1984) 45.

\bibitem{WilczekZee1984}
F.~Wilczek and A.~Zee,
``Appearance of gauge structure in simple dynamical systems,''
Phys.\ Rev.\ Lett.\ \textbf{52} (1984) 2111.

\bibitem{AiCruzGarbrechtTamarit2022}
W.-Y.~Ai, J.~M.~Cruz, B.~Garbrecht and C.~Tamarit,
``The limits of the strong CP problem,''
JHEP \textbf{06} (2022) 148,
arXiv:2205.15093 [hep-ph].

\bibitem{AiGarbrechtTamarit2024}
W.-Y.~Ai, B.~Garbrecht and C.~Tamarit,
``CP conservation in QCD as a consequence of the order of limits,''
arXiv:2404.16026 [hep-ph].

\bibitem{AlbandeaCatumbaRamosPRD}
D.~Albandea, F.~Catumba and R.~O.~Ramos,
``Topological susceptibility and the strong CP problem in the quantum rotor,''
Phys.\ Rev.\ D \textbf{110} (2024) 094512,
arXiv:2402.17518 [hep-th].

\bibitem{Benabou:2025viy}
J.~N.~Benabou, A.~Hook, C.~A.~Manzari, H.~Murayama and B.~R.~Safdi,
``Clearing up the Strong $CP$ problem,''
arXiv:2510.18951 [hep-ph].

\bibitem{Khoze:2025auv}
V.~V.~Khoze,
``A note on instantons, $\theta$-dependence and strong CP,''
arXiv:2512.06827 [hep-ph].

\bibitem{Sannino:2026wgx}
F.~Sannino,
[arXiv:2601.19735 [hep-ph]].

\bibitem{Aghaie:2026pkf}
M.~Aghaie and R.~Sato,
[arXiv:2601.18248 [hep-ph]].

\bibitem{Ringwald:2026apz}
A.~Ringwald,
[arXiv:2601.04718 [hep-ph]].

\end{thebibliography}
\end{document}